# Compact and accurate chemical mechanism for methane pyrolysis with PAH growth

A. Khrabry[1], I.D. Kaganovich[2], Y. Barsukov[2], S. Raman[3], E. Turkoz[3], D. Graves[1]

[1]Princeton University, Princeton, NJ, USA

[2]Princeton Plasma Physics Laboratory, Princeton, NJ, USA

[3]ExxonMobil Technology and Engineering, Annandale, NJ, USA

**Abstract**

A reliable and compact chemical mechanism of gas-phase methane pyrolysis leading to formation of large polycyclic aromatic hydrocarbon (PAH) molecules has been developed. This model is designed for studies of carbon nanostructure synthesis such as carbon black and graphene flakes, including soot growth kinetics. Methane pyrolysis with carbon nanostructure synthesis is a two-stage process where conversion of $CH_4$ to $C_2H_2$ precedes the growth of PAH molecules from acetylene. We present a single chemical mechanism that accurately describes both stages. We have constructed a compact and accurate chemical mechanism capable of modeling both stages of methane pyrolysis based on the ABF[1] mechanism which was expanded with most prominent reaction pathways from the mechanism by Tao[2] for small PAH molecules and HACA pathways for larger PAH molecules, up to 37 aromatic rings. The resulting mechanism was validated through comparison to multiple available sets of experimental data. Good agreement with the experimental data for both processes was obtained. Performance of the mechanism was tested for pyrolysis of methane-rich mixtures under long residence times leading to abundant formation of PAH molecules. It is shown that the inclusion of larger PAH species (up to A37) in the chemical mechanism is important for accurate prediction of the fraction of carbon converted to PAH molecules and, correspondingly, residual fraction of acetylene in the mixture.

1. Introduction

Pyrolysis of natural gas (methane) is a promising way to avoid $CO_2$ emissions associated with traditional production of hydrogen through steam reforming of methane[3,4,5,6,7,8,9,10]. The process can be made commercially viable by producing carbon nanostructured by-products in one or another valuable form which could include carbon black[11,12] (CB), carbon nanotubes (CNTs), graphene flakes (GFs), etc. Industrial-scale non-catalytic plasma pyrolysis of methane to hydrogen and CB has been achieved, e.g., by Monolith Materials[3]. Industrial-scale synthesis of CNTs in hydrocarbon pyrolysis has also been achieved by e.g., OCSiAl[13,14]. However, despite recent advances in synthesis of carbon nanostructures from methane, the cost of the materials produced is still relatively high. There is a need for better understanding of the mechanism of gas phase pyrolysis chemistry to improve the processes used for amorphous carbon and nanostructured formation. An improved model will help develop an optimized reactor design and process conditions for the effective synthesis of carbon nanostructures.

Potential nanostructures of interest include CB, CNTs, GFs and soot, with the first three being the desired products and the latter is generally undesired. Soot formation is detrimental for the synthesis process as it consumes valuable carbon feedstock and contaminates the product but is often difficult to avoid for CNT synthesis while maintaining high CNT yield[15]. While CNTs require catalytic nanoparticles for their growth, CB, GFs, and soot grow non-catalytically in the gas phase from polycyclic aromatic hydrocarbon (PAH) precursor molecules[16]. Several models have been developed recently describing formation and growth of CB and soot particles from PAH molecules, e.g., Refs. [17, 18, 19, 20, 21, 22, 23].

The foundation for these models is a gas phase chemical kinetic mechanism that describes formation of PAH molecules that are precursors for insipient primary particles within CB, soot and GFs. As was shown in Ref. [24], the precursor species for the PAH molecules is acetylene ($C_2H_2$). The growth of PAH molecules in the pyrolysis of methane is preceded by conversion of $CH_4$ to $C_2H_2$. Multiple chemical mechanisms have been reported in the literature that have been tested separately for: 1) high temperature pyrolysis of $CH_4$ to form $H_2$ and $C_2H_2$, and 2) PAH growth from $C_2H_2$ mixtures in the oxygen-free environment at longer residence times. Among these mechanisms, there are mechanisms that demonstrated good agreement with experimental data for one or another stage, e.g., the ABF[1] mechanism has been demonstrated to work accurately for stage 1 (pyrolysis of methane to $C_2H_2$) in Refs. [25,[26]], and the Tao[2] mechanism works well for Stage 2. However, to model accurately the formation of carbon nanostructures in methane pyrolysis, a single mechanism is needed for both stages of the process. To the best of our knowledge, no such mechanism has been developed and/or tested for both stages. Furthermore, the Tao mechanism is quite large; it has hundreds of species and thousands of reactions, most of which likely play minimal roles. We therefore aimed to develop and test a more compact mechanism covering only the most critical pathways for PAH growth.

We note the need for such a model to include PAH molecules up to a size which is typical for primary CB/soot particles. This would be PAH molecules of about ~2-3 nm in diameter[18,27,28,29,30] having on the order of 100 carbon atoms[28]. This corresponds to ~20-50 aromatic rings. Molecular dynamics (MD) modeling of carbon nanostructures[31] has shown that PAH molecules smaller than circumcoronene ($C_{54}H_{18}$) do not adhere sufficiently to form insipient soot/CB particles[31] at typical temperatures for soot growth, i.e., 1000 K–1500 K[32]. The Tao mechanism[2] accounts for relatively small PAH molecules only, up to five aromatic and/or 5-membered rings. One motivation for the present model is to incorporate larger PAH molecules in the mechanism. Fortunately, abundant thermodynamic data from quantum chemical calculations for various PAH species up to ovalene ($C_{32}H_{14}$) [33] is now available that allows expansion of the Tao mechanism[2] based on the broadly accepted dominant pathway for PAH growth – the hydrogen-addition-carbon-abstraction mechanism (HACA)[24,34,35,36,37,38,39].

In the present study, we analyze the available chemical mechanisms and derive a relatively compact extension of the ABF[1] mechanism with selected chemical reactions from the Tao[2] mechanism. Additional PAH species (up to 37 aromatic rings) are included such that both stages of methane pyrolysis can be combined, leading to accurate predictions of the formation kinetics of realistic carbon nanostructures.

2. Development of the chemical mechanism for pyrolysis of methane

Earlier studies have shown the ABF mechanism[1], originally designed to model PAH formation in combustion processes, actually works quite well for the first stage of methane pyrolysis, i.e., transformation of $CH_4$ into $C_2H_2$ [25]. We have performed extensive testing of the ABF mechanism in comparison to other available mechanisms for a broad range of experimental conditions corresponding to various temperatures and methane dilution ratios with different gases. Results of these tests are presented in Appendix 3. We conclude that the ABF mechanism is generally superior to other published mechanisms. It produces results in good agreement with experimental data in the entire range of the conditions tested whereas other mechanisms either performed well in some conditions but poorly in other conditions or produced results of generally limited accuracy for all conditions considered. The ABF mechanism is also rather compact: it has only about 70 species and 240 reactions for non-oxygen species.

However, more recent studies, such as that by Tao[2], have shown that the performance of the ABF mechanism at predicting the growth of PAH molecules has limitations. Although the ABF mechanism is one of the pioneering mechanisms considering the growth of PAH molecules, it only considers small PAH molecules up to A4 (four aromatic rings). It predominantly includes chemical pathways based on the HACA and HAVA[37,39] pathways. However, later experimental data[40,41,42,43] and further development of the PAH growth mechanisms have shown that the ABF mechanism is not as accurate as later mechanisms. There appear to be other chemical pathways that are important for PAH growth and need to be included in the mechanism.

On the other hand, the mechanism by Tao[2] was specifically designed to model formation of small PAH species which have up to five aromatic/five-membered rings. It is in essence an assembly of chemical pathways discovered earlier and included in previous chemical mechanisms[44,45,46,47]. The mechanism has been validated in Ref. [2] by comparing to available experimental data on the growth of small PAH species during pyrolysis of $C_2H_2$. The resulting mechanism has demonstrated a good agreement with the experimental data on concentrations of selected PAH species as compared with previous mechanisms. However, even though this mechanism includes chemical pathways of the first step of methane pyrolysis, i.e., transformation of $CH_4$ to $C_2H_2$, the performance of the mechanism has not been tested or validated for this process. Also, the mechanism includes about 300 species and more than a thousand of chemical reactions, and this size limits its practical utility.

### 2.1. Expanding the ABF mechanism with most crucial PAH formation pathways from the Tao mechanism

We analyzed the ABF and Tao mechanisms and determined key pathways for PAH formation during pyrolysis of $C_2H_2$ that were present in the Tao mechanism but were missing from the ABF mechanism. Cantera's in-built pathway visualization tool was used for this analysis. These crucial pathways are presented in Fig. 1. These pathways were then included in our version of the expanded ABF mechanism. Performance of the expanded and original ABF mechanisms and the Tao mechanism is compared in subsequent sections of the paper. Rate constants for the reactions that are added to our expanded ABF mechanism are summarized in Table 1 of Appendix 2 along with literature sources.

Crucial pathways of formation of $C_6H_6$ (A1, benzene) are shown in Fig. 1a. Benzene (A1) is one of the major intermediate species in the pathways to formation of larger PAH species (A2, A3, A4 etc.), and therefore these pathways not only affect the production rate and concentration of A1 but, in fact, production rates and concentrations of all PAH species. The rate constant of the reaction of $C_4H_2$ formation from $C_2H_2$ originates from Ref. [48] and is based on the analysis of experimental data. The rate constants of the reactions of $C_4H_2$ transformation to $H_2CCCCH$ and its further transformation to $C_6H_5$ are taken from Ref. [49]. The importance of this pathway for the formation of the first aromatic ring (i.e., benzene), has already been noted in Ref. [50] where chemical mechanisms of benzene formation were validated by comparing to available experimental data. The rate constant of $C_6H_6$ formation from $C_4H_4$ and $C_2H_2$ originates from Ref. [51] and is based on the analysis of experimental data.

Crucial pathways of destruction of A2R5 (acenaphthylene) and A3 (phenanthrene) are shown in Fig. 1b (pathways of formation of A2R5 and A3 are similar in the ABF and Tao mechanisms). These A2R5 destruction pathways are in fact the established pathways of the HACA mechanism where two acetylene molecules sequentially attach to the PAH molecule leading to a formation of an additional aromatic ring while the existing 5-membered stays in place. This results in the formation of either A3R5 or A3LR5. The additional mechanism of A3 destruction is somewhat similar to the HACA mechanism with the difference that here a $C_2H_2$ molecule attaches to the zig-zag edge of an A3 molecule leading to a formation of a 5-menbered ring. Growth pathways of PAH molecules with 5-membered rings are not typically included in chemical mechanisms PAH growth because the presence of a 5-membered ring limits the possibilities of the molecule growth. Once the 5-membered ring forms it 'locks' the edge of the molecule as it stays in place and prevents further growth of the molecule in this direction through HACA mechanism. However, the inclusion of these particular pathways of A2R5 and A3 destruction is important as it substantially reduces resulting concentrations of A2R5 and A3 molecules (which are among dominant species in the early stages of soot formation) and thereby plays an important role for the determination of the total rates of $C_2H_2$ consumption and larger PAH formation.

A crucial pathway for the formation of A4 (pyrene) is shown in Fig. 1c. We note that this additional pathway has nothing to do with the HACA pathway (HACA pathway of A4 formation was already included in the ABF mechanism). Our chemical pathway analysis has shown that this additional pathway in fact plays a dominant role in the formation of A4. Interestingly, this pathway seems to be applicable merely to formation of A4 and cannot be expanded to larger PAH molecules unlike the HACA pathway because this pathway is built upon particular intermediate species, namely, $C_7H_7$ (benzyl radical) which forms from two linear-shaped species $C_4H_4$ and L-$C_3H_3$. Benzyl radical is further transformed to $C_9H_8$ (indene) and $C_9H_7$ (indenyl) through reactions similar to those from the HACA mechanisms. Subsequently, benzyl and indenyl radicals merge to form A4. The rate of the latter reaction originated from Ref. [52] where it was estimated using similarity to the reaction of A2 (naphthalene) formation from merging of two $C_5H_5$ (cyclopentadienyl) radical molecules (2 $C_5H_5$ <=> A2 + $H_2$). The rate constant for the latter reaction was determined in Ref. [53] by applying the quantum Rice-Ramsperger-Kassel (QRRK) formalism which is based on the transition-state theory. Notoriously, the rate constant of this reaction was adopted to the Tao mechanism with a typo: the constant was accidentally increased by a factor of 10 ($2\times10^{12}$ was used instead of $2\times10^{11}$ in the pre-exponential factor of Arrhenius formula). Fortunately, the effect of this rate constant on the production rate of A4 is rather small because the rate-limiting step in the A4 formation pathway is the formation of benzyl radical, not the merging of benzyl and indenyl. We corrected this typo and implemented the original rate from Ref. [52] to the expanded ABF mechanism. Rate constants for the reaction of $C_7H_7$ formation was computed in Ref. [54] for a reversed reaction using ab initio quantum chemical modeling.

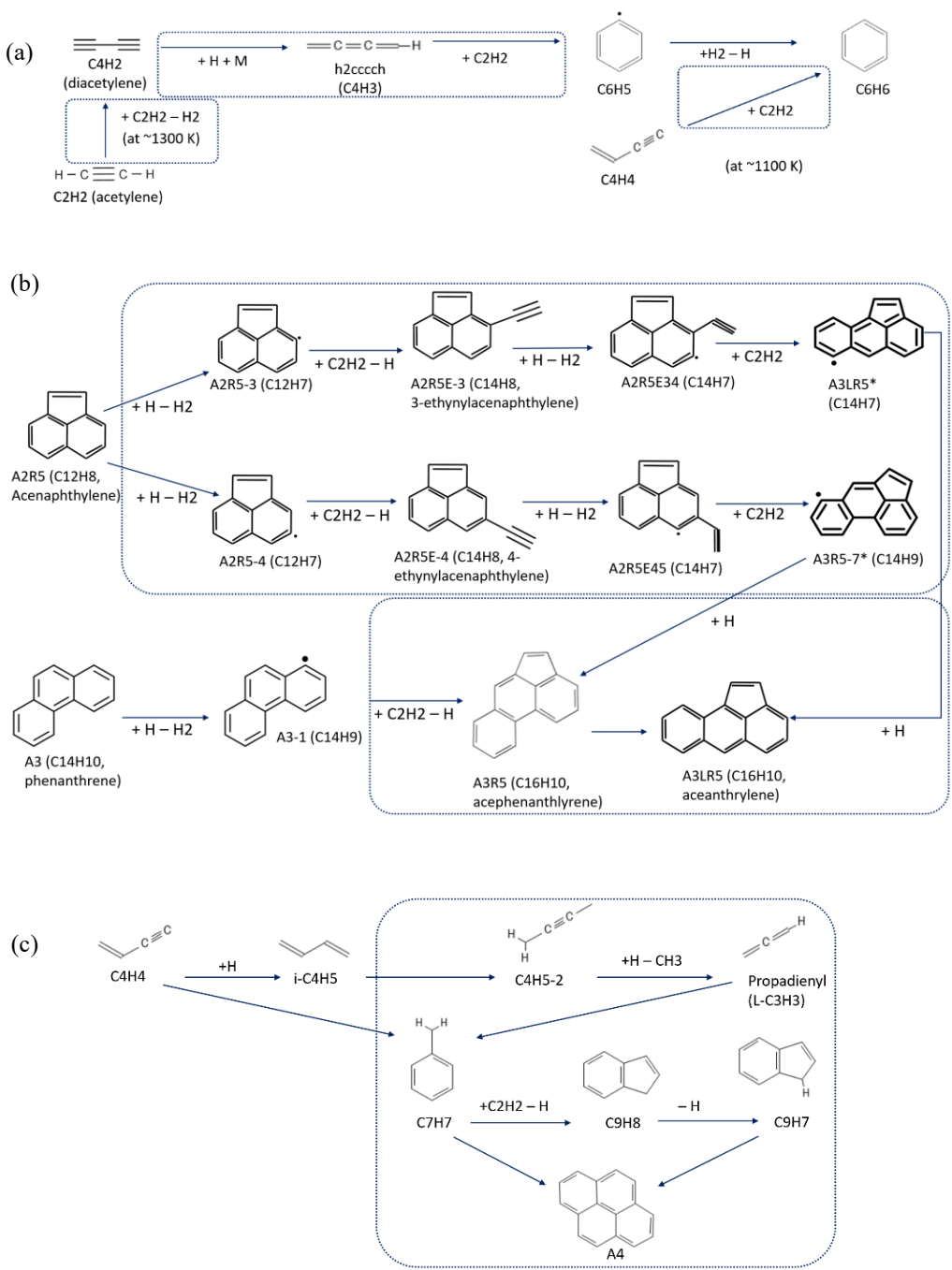

Fig. 1. Crucial pathways of PAH growth that are present in the Tao mechanism[2] but are missing in the original ABF[1]. Pathways of A1 formation (a), A2R5 and A3 destruction (b), and A4 formation (c). Species and reactions missing in the original ABF mechanism are highlighted by the dotted frames.

### 2.2. Expanding the mechanism further to include larger PAH species up to A37

In addition to expanding the ABF[1] mechanism with selected chemical reactions from the Tao[2] mechanism which are responsible for accurate prediction of formation of small PAH species (up to A4), we expanded the mechanism further to include larger PAH species up to A37 which are thought to be precursors for insipient particles in CB/soot. This expansion was made based on the broadly accepted HACA mechanism[24,37] of PAH growth and new thermodynamic data on all kinds of PAH species up to A10 (ovalene) [33] which had been derived from quantum chemistry modeling carried out using the density functional theory (DFT) method which is known to produce reliable optimized geometries and thermodynamic properties of molecules.

We performed the further mechanism expansion in two steps. As the first step, we used the thermodynamic data[33] to determine the major pathway of PAH growth from A4 to A10. According to the HACA mechanism, the growth of PAH species proceeds through sequential addition of $C_2H_2$ molecules leading to an incremental increase in the number of aromatic rings in a PAH molecule (see Figs. 2 and 3). A new aromatic ring can form either on the corner of a PAH molecule, as shown, e.g., for a transition from A10 to A11- on Fig 3, or in a "bay" region of an armchair edge of a PAH molecule, as shown, e.g., for transitions from A11 to A12, A12 to A13 and A13 to A14 on Fig 3. A new ring cannot form on a zig-zag edge of a molecule through this mechanism. This limits the number of possible pathways to form A10 from A4 through HACA mechanism. We assembled a list of possible pathways to form A10 from A4 through PAH species with stepwise addition of aromatic rings in accordance with the HACA rules. The thermodynamic data was used to tule out pathways involving least stable PAH species and come up with a graph of pathways shown on Fig. 2. We added these new species and chemical reactions from these pathways to the ABF mechanism. Rate constants for the forward reactions were adopted from the original HACA mechanism[24,37]. Rates of reverse reactions were calculated from the principle of detailed balance using the thermodynamic data[33].

We tested the resulting mechanism with PAH species up to A10 for the conditions of experiments[42] and found out that in fact only one branch (one pathway) of A10 formation completely dominated over other pathways. This is the pathway that involves coronene (the pathway highlighted by a dashed frame on Fig. 2). Other pathways could be safely removed from the mechanism with no effect on the concentration and production rate of coronene. It was also observed that concentrations of species which have low symmetry due to "protruding" aromatic rings such as benzo[e]pyrene, dibenzo[b,ghi]perylene, benzo[a]coronene etc. have much lower concentrations in the resulting mixture than other species, whereas symmetric species such as coronene and ovalene have the highest concentrations. This is not surprising that most symmetric species which have the largest number of carbon-carbon bonds among molecules with the same number of carbon atoms appear to be the most stable due to higher number of C-C bonds and increased delocalization and/or resonating structures Accordingly, the dominant PAH growth pathway tends to pass through symmetric, convex PAH molecules. We used this knowledge in the second step of the mechanism development. We extrapolated the thermodynamic data from Ref. [33] to PAH molecules beyond ovalene based on the number of C-C and C-H bonds in a molecule. We constructed the mechanism so that it passes through all symmetric convex PAH molecules on the way to A37 as shown on Fig. 3 with a detailed list of corresponding HACA reactions. As soon as a symmetric PAH molecule forms, the next PAH molecule will have an additional aromatic ring "sticking out" at the corner. This results in two adjacent armchair "bays" at two adjacent edges. Once a bay gets "filled" by a new aromatic ring, a new adjacent "bay" forms, and so on, until the entire edge is "filled". Once both edges are "filled", a new symmetric PAH molecule is obtained. The mechanism can be expanded to arbitrarily large PAH molecules using this procedure. We decided to stop at A37 which seems to be a reasonable size for soot precursors. We discuss the results of applying the original and expanded ABF mechanism in the next section.

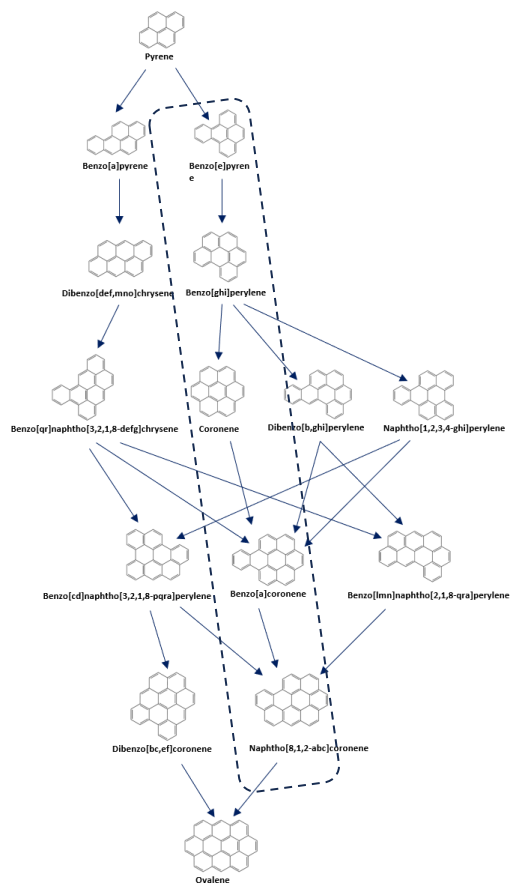

Fig. 2. Chemical pathways of PAH molecules growth from pyrene (A4) up to ovalene (A10).

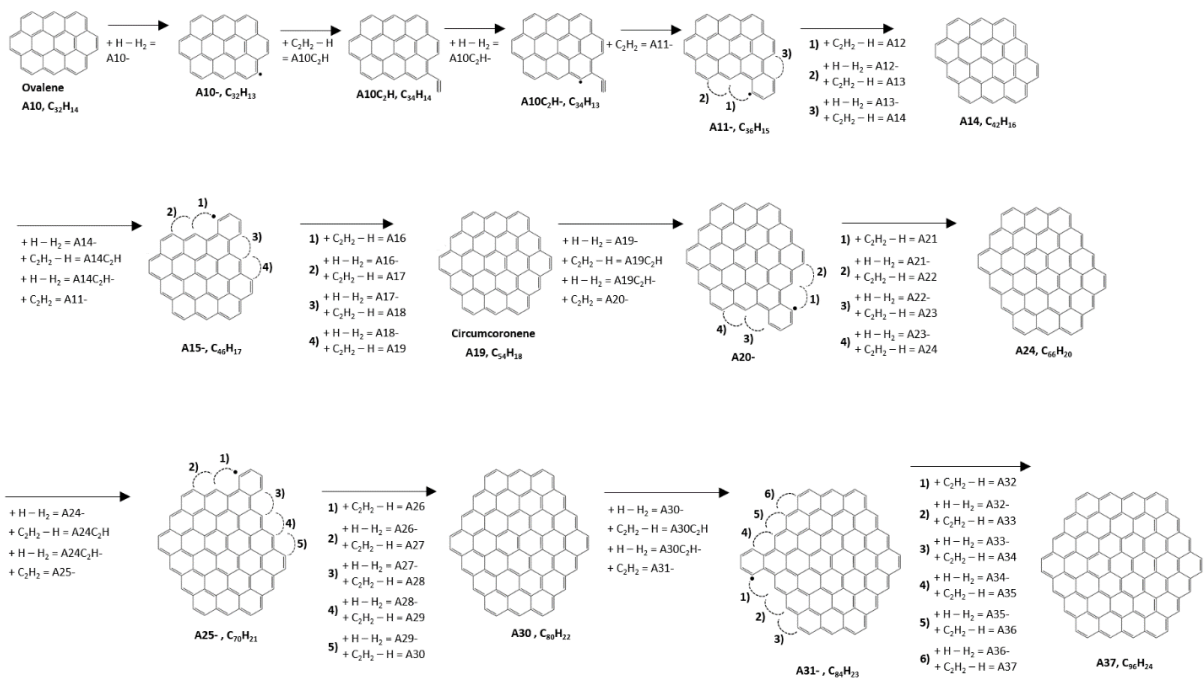

Fig. 3. Chemical pathway of PAH molecules growth from Ovalene (A10) to A37.

## 3. Validation of the proposed methane pyrolysis growth mechanism: comparing to ABF[1] and Tao[2] mechanisms for a broad range of conditions

### 3.1. Testing the proposed mechanism for methane transformation to acetylene

In this section, we compare the performance of the original ABF[1] and Tao[2] mechanisms to the proposed mechanism in application to the first stage of methane pyrolysis (its transformation to $C_2H_2$). The proposed mechanism is in essence the ABF mechanism expanded with selected chemical reactions from the Tao mechanism and additional reactions to account for the formation of larger PAH molecules (up to A37). More detailed comparison of these mechanisms to other mechanisms available in literature for pyrolysis of methane into acetylene can be found in Appendix 3 where it is shown that the original and expanded ABF mechanisms outperform other mechanisms in the totality of the tests. Note that such a thorough study of ABF mechanism's performance has not been done before. The mechanisms have been tested for a broad range of conditions for which experimental data was available in literature. These include various temperatures, $CH_4$ dilution ratios and background gases. All results in details are presented in Appendix 3. Here, we present selected results to outline the scope of the tests and derive important conclusions on the performance of the original ABF and Tao mechanisms in comparison to the proposed mechanism.

In Fig. 4, modeling results for pyrolysis of methane modestly diluted with hydrogen ($H_2$:$CH_4$ = 2:1) are presented in comparison to experimental data from Ref. [55]. In these experiments, a tubular flow reactor was used which was equipped with a cold finger that could be moved along the reactor tube to quench the process and effectively control residence time. $CH_4$ conversion degree as a function of residence time was measured. Temperature in the heated zone of the reactor was constant. The heated zone (constant temperature zone) was preceded by a short preheating zone which was modeled by a linear increase of temperature during the first 0.15 s. At the conditions and residence times of the experiments, soot formation was not observed, i.e., the experiments represented solely the first stage of methane pyrolysis, i.e., transformation of $CH_4$ to $C_2H_2$. The results in Fig. 4 are presented for two heated zone temperatures 1200 C and 1400 C are presented which are within the temperature range of interest in which carbon nanostructures would typically grow during the second stage of the pyrolysis process (if residence time was longer). The plots show how methane conversion progresses with time. Apparently, the process is much faster at higher temperatures.

As is evident from the figure, the original and expanded ABF mechanisms produce the results that are very close to each other confirming that PAH formation is moderate at these residence times as it has a low effect on overall conversion of methane (the effect of PAH formation will be discussed in more detail in the next section). The results of the ABF mechanisms are very close to the experimental data. On the contrary, the Tao mechanism overpredicts multifold the rate of methane conversion leading to a much higher conversion degree. This discrepancy with the experimental data is even larger at a lower temperature. In Fig. 5, for comparison, the results for considerably lower temperature of 765 C are shown. Experimental data is taken from the experiments[56], where pure methane was gradually decomposed in a static chamber reactor at a constant low temperature. Mole fraction of $H_2$ that was released from methane decomposition is plotted as a function of residence time. The process is slow at this temperature, hence only a sub-percent fraction of $CH_4$ is decomposed during the observation period of 1000 s. The original and expanded ABF mechanisms produced almost identical results in terms of $H_2$ mole fraction, in good agreement with the experimental data. On the other hand, the Tao mechanism drastically overpredicted the process rate: corresponding line exhibits almost vertical growth on the chart.

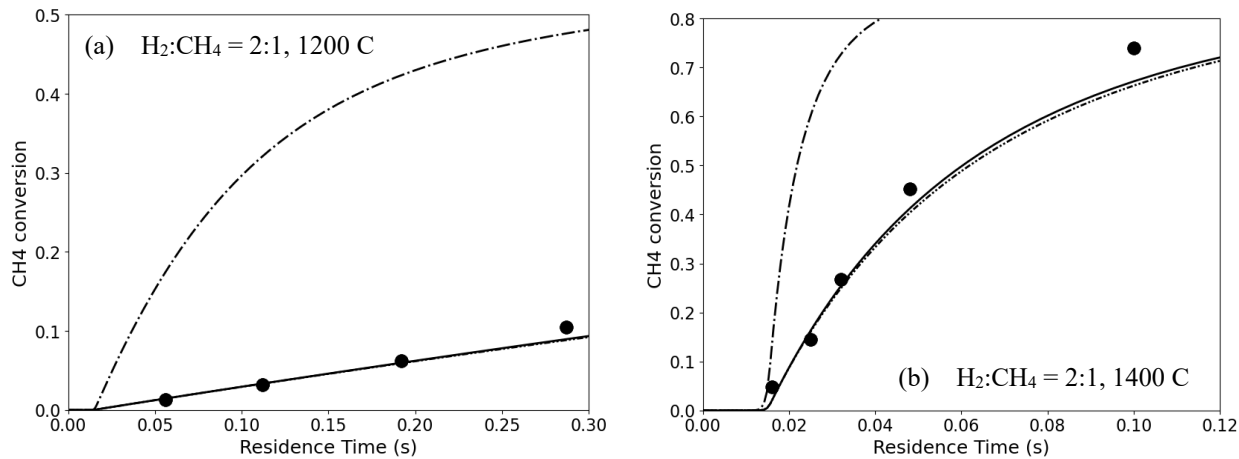

Fig. 4. Methane conversion degree as a function of residence time computed using various chemical mechanisms (lines of different styles) compared to the experimental data in decomposition of methane diluted with hydrogen $H_2:CH_4 = 2:1$ at 1200 C (a) and 1400 C (b). ▬▪▪▬▪▪ ABF mechanism (original); ▬▪▬ the Tao mechanism (2019); ▬▬ the proposed mechanism (i.e., ABF mechanism added with the Tao mechanism reactions and expanded with larger PAH molecules); ● experimental data [55].

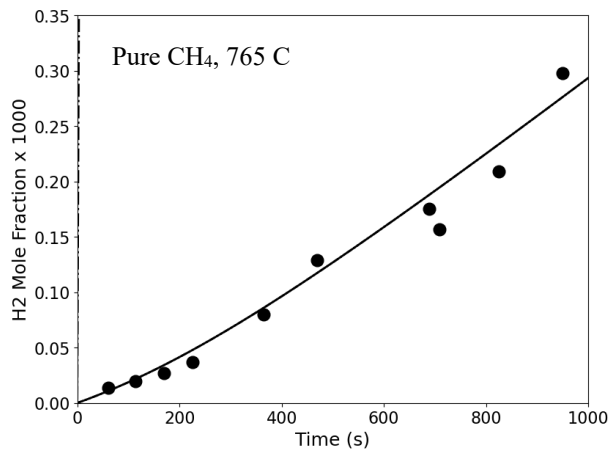

Fig. 5. Mole fraction of $H_2$ as a function of residence time in decomposition of pure methane at 765 C. Line styles are the same as in Fig. 4. Markers - experimental data[56].

Comparisons of the modeling results to the experimental data, for decomposition of methane strongly diluted in helium[57] and nitrogen[58] are presented in Fig. 6. In both experiments, a tubular flow reactor with a constant-temperature heated zone was used. Temperature was varied in a range ~1000 C – 1200 C in a series of experiments. Molar fraction of $CH_4$ at the outlet was measured for each reactor temperature. An isothermal plug-flow reactor model was used for simplicity to model these experiments with residence time defined as a function of temperature: 7550K/T (s) and 4550K/T (s) in the cases of helium and nitrogen respectively (residence time is inversely proportional to temperature to account for thermal expansion of the gas in a constant flow and constant length reactor). Final mole fraction of $CH_4$ over is initial fraction is plotted in Fig. 6 as a function of temperature.

Similar to the previous experimental conditions considered, the original and expanded ABF mechanisms are in an excellent agreement with the experimental data for the $CH_4$-He mixture (Fig. 6a). The agreement is satisfactory for the $CH_4$-$N_2$ mixture. Again, the Tao mechanism overpredicts the rate of the process resulting in underestimated $CH_4$ mole fraction at the reactor's outlet.

To summarize this section, the original ABF mechanism as well as the proposed expanded version of the ABF mechanism outperform the Tao mechanism when applied to the first stage of methane pyrolysis (i.e., conversion of $CH_4$ to $C_2H_2$). The results are in a good agreement with the experimental data for the entire range of conditions considered. By contrast, the Tao mechanism considerably overpredicts the rate of methane decomposition, with strong deviation from the experimental data at low temperatures.

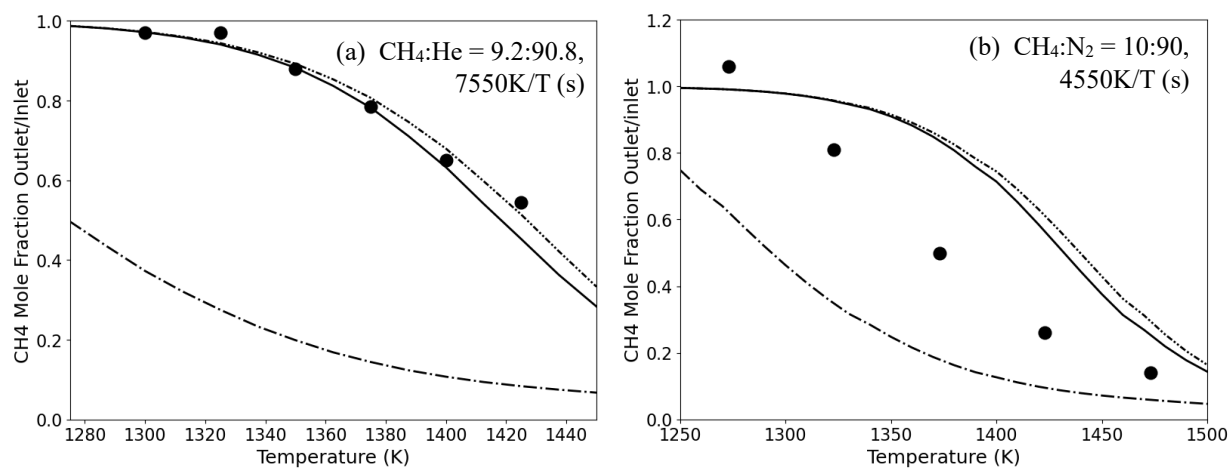

Fig. 6. Final mole fraction of CH4 over its initial fraction at the entrance of the flow reactor as a function of temperature computed using various chemical mechanisms (lines of different styles, as in Fig. 4) compared to the experimental data from Refs. [57] (a) and [58] (b) – circles.

### 3.2. Testing the proposed mechanism in the second stage of methane pyrolysis: PAH growth from acetylene

In this section, we compare performance of the original ABF mechanism and the proposed mechanism to the Tao mechanism in application to the conditions typical for the second step of methane pyrolysis, i.e., conversion of $C_2H_2$ to PAH molecules. Experimental data on the growth of PAH molecules is taken from Ref. [42]. In these experiments, a mixture of $C_2H_2$ highly diluted with $N_2$ (3% $C_2H_2$ – 97% $N_2$) was passed through a heated-wall tubular flow reactor with the temperature varying within the range of intense soot growth, i.e., 1000 K–1500 K). Densities of small PAH species were measured at the outlet of the reactor. We modeled this reactor using a plug-flow model with a temperature profile resembling both the heating zone and preset temperature zone of the reactor as described in Appendix 1. Residence time in the preset temperature zone was inversely proportional to the preset reactor temperature, to account for thermal expansion of the gas, as it was done in earlier modeling[2].

Final concentrations of small PAH species (A2, A3 and A4) as well as A2R5 as functions of the reactor preset temperature are plotted in Fig. 7 in comparison with the experimental data. $C_2H_2$ conversion degree and $H_2$ yield as functions of the reactor temperature are shown in Fig. 8. Residence time at each preset reactor temperature was defined as 1706 K/T (s). We also performed additional tests with extended residence time defined as 4552 K/T (s) for which experimental data was available as well. The results for the extended residence time were similar and can be found in Appendix 4. We used two versions of the expanded ABF mechanism in the modeling. The first one (denoted as ABF+Tao, selected) was expanded with selected reactions from the Tao mechanism to improve modeling of small PAH molecules. The second one (denoted as ABF+Tao, selected + large PAH) was additionally expanded with a HACA pathway to include formation of larger PAH species up to A37.

As can be seen from these figures, the original ABF mechanism clearly lacks important pathways responsible for the transformation of $C_2H_2$ to PAH species. It underestimates final concentrations for all PAH species under consideration by orders of magnitude at temperatures below 1300K, for both residence times. $C_2H_2$ conversion degree and $H_2$ yield are also noticeably underestimated by the original ABF mechanism as well, for all temperatures considered. On the contrary, the predictions of the Tao mechanism which was specifically designed to model formation of small OAH molecules are in a good agreement with the experimental data. The ABF mechanism expanded with selected chemical reactions from the Tao mechanism (as described in the precious sections and shown in Fig. 1) exhibits drastically better behavior than the original one: modeling results are much closer to the experimental data, especially in the intermediate temperature range.

Further expansion of the mechanism with larger PAH species (up to A37) does not have any noticeable effect on the concentrations of A2, A3 and A2R5 as well as $C_2H_2$ conversion rate and $H_2$ yield. The lines corresponding to the mechanism expanded with selected reactions from the Tao mechanism and further expanded with larger PAH molecules virtually coincide on the plots. However, the concentration of A4 molecules is noticeably reduced at higher temperatures when larger PAH molecules are included in the mechanism resulting in considerably better agreement with the experimental data. This result is not surprising since the improvement of the mechanism is, in essence, realized through a pathway of A4 transformation to larger PAH molecules.

The inclusion of pathways of A2R5 and A3 destruction shown in Fig. 1b is important as it substantially reduces concentrations of these species and brings the results closer to the experimental values, though there is still some overprediction in the results. The inclusion of multiple other minor pathways of A2R5 and A3 destruction from the Tao mechanism could make the agreement with the experimental data slightly better. However, this study aims to focus on deriving a compact mechanism that includes both pyrolysis of $CH_4$, $C_2H_2$ and mixtures of $CH_4$ and $C_2H_2$. Also, some of the deviation from the experimental data might be attributed to the limitations of the 1D plug-flow reactor model used here and in previous papers. A more accurate study of 2D reactor effects is a subject of future studies for which this compact mechanism can be conveniently utilized.

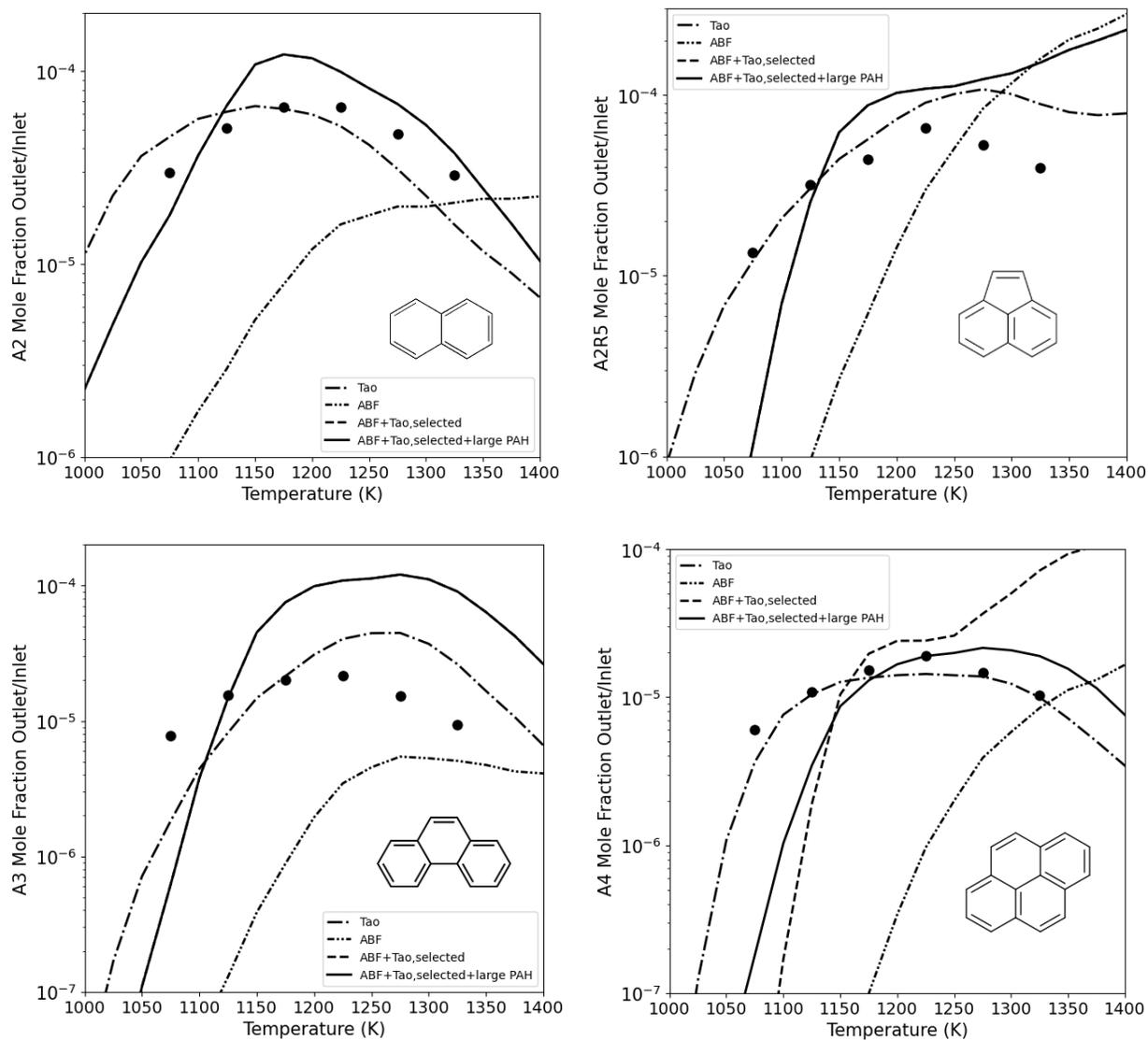

Fig. 7. Computed final concentrations of various PAH species at an outlet of a constant-temperature flow from the experiments[42] after the residence time of 1706 K/T (s), i.e., approx. ~1.5s. ⋯⋯ ABF mechanism (original); --- ABF mechanism added with selected reactions from the Tao mechanism; —— ABF mechanism added with the Tao mechanism reactions and expanded with larger PAH molecules, —·— the Tao mechanism; ● experimental data[42].

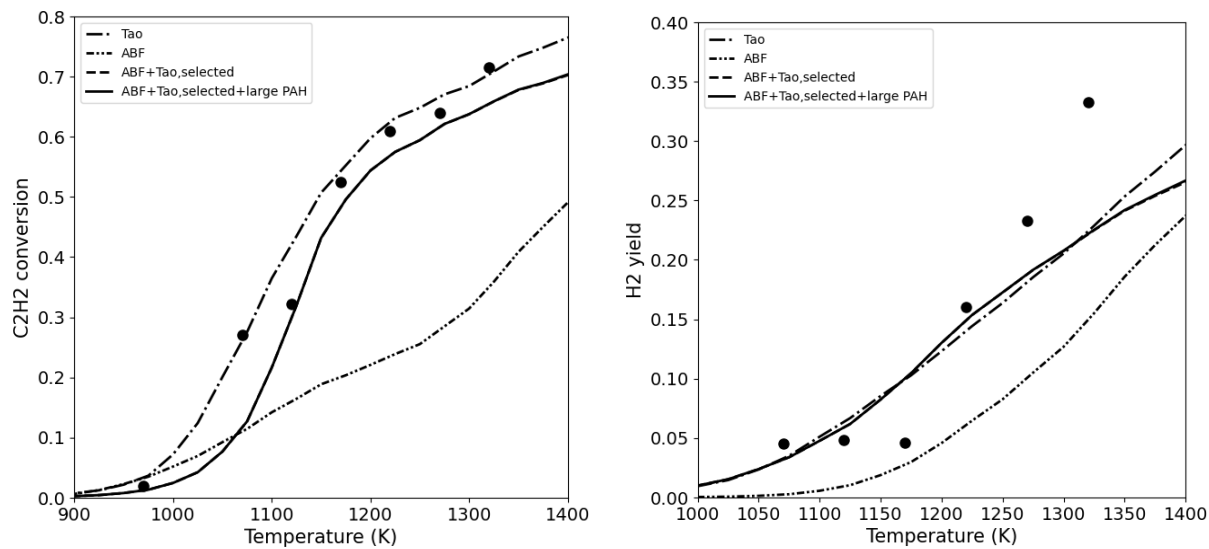

Fig. 8. $C_2H_2$ conversion degree and $H_2$ yield after the residence time of 1706 K/T (s). Line styles and markers have the same meaning as on Fig. 7.

### 4. Applying the expanded ABF mechanism to model pyrolysis of $CH_4$ and $C_2H_2$

In this section, we apply the improved ABF mechanism[1] with selected (most prominent) reaction pathways from the Tao mechanism[2] and HACA pathways for larger PAH species up to A37 to perform modeling studies of methane pyrolysis. A plug-flow model of an isothermal reactor is used. We consider temperatures covering the typical range of PAH growth, i.e., from 1100 C to 1600 C. Two initial gas compositions were modeled: pure methane and a slightly diluted methane-hydrogen mixture ($CH_4$:$H_2$ = 1:1) to facilitate PAH growth in methane-rich environment. Long residence time of 5 s up to 100 s was modeled to cover both stages of the process: the faster first stage of $CH_4$ transformation to $C_2H_2$ and the slower second stage of PAH formation and growth from $C_2H_2$.

Modeling results are presented in Figs. 9 and 10 for pure and slightly diluted methane mixtures respectively. Fractions of carbon in major species are plotted as a function of residence time. All PAH species from A2 to A37 are combined and shown by a single black line. A1 and A2R5 are presented by separate lines. We omit secondary species from the charts such as A3R5, A3LR5, $C_4H_2$, $C_4H_4$ and $C_6H_2$ to avoid overcomplication of the chart. A3R5 and A3LR5 typically have considerably lower concentrations than A2R5, and $C_4H_2$, $C_4H_4$ and $C_6H_2$ have much lower concentrations than $C_2H_2$. In order to assess the extent at which large PAH species form and the role they play, we compare the results obtained using three versions of the ABF mechanism represented by different line styles. Solid lines correspond to the ABF mechanism expanded with selected reactions from the Tao mechanism (only small PAH species up to A4 are included). Dashed lines correspond to the mechanism additionally expanded with PAH species up to ovalene as shown in Fig. 2. Dotted lines correspond to the mechanism additionally expanded with PAH species up to A37 as shown in Fig. 3.

As is evident from the figures, at these conditions, $CH_4$ transforms to $C_2H_2$ quite quickly (within a sub-second range for T ≥ 1300 C and within several seconds for T = 1100 C). At this point in time, $C_2H_2$ concentration peaks and starts to decline as PAH species form and start to grow and consume carbon. This qualitative behavior is captured by all versions of the chemical mechanism used. The concentration of benzene (A1) peaks almost simultaneously with $C_2H_2$. Later in time, they are consumed to form PAH species. The concentration of A2R5 reaches its peak at a later time, and the total amount of PAH species monotonically grows towards its thermodynamic limit. The PAH growth is a relatively slow process which happens within a time frame of on the order of seconds even at the highest temperature considered.

In qualitative agreement with previous experiments[59,60,61], our model shows that PAH molecules form more abundantly in the case of pure methane feedstock. In this case, up to 70%-90% of carbon ends up in PAH species in a broad temperature range from 1300 C to 1500 C. For comparison, in the case of methane diluted with hydrogen at 1:1 molar ratio, the conversion of $C_2H_2$ to PAH of ~80% is achieved only within a narrower temperature range close to 1400 C. In all the conditions considered, the growth of PAH species expanded beyond four aromatic rings (i.e., beyond pyrene) as it is manifested by larger total amount of carbon accumulated in PAH molecules when the mechanisms expanded to larger PAH species are used. At temperatures ≥ 1400 C the amount of carbon in larger PAH molecules is on order of tens of percent; the growth of PAH molecules expands beyond ovalene (A10) as follows from the fact that the largest mechanism which accounts for PAH species up to A37 predicts larger amount of PAH production than the mechanism which account for PAH molecules up to ovalene only.

At lower temperatures, the formation of PAH molecules occurs at a considerably slower rate. The range on the x-axis is expanded to 100 s and 30 s for temperatures of 1100 C to 1200 C, respectively, to illustrate the growth process. PAH species don't grow beyond ovalene (carbon fractions in species A2-A10 is the same as in all species A2-A37). Transformation of $CH_4$ becomes less efficient at these temperatures. About 50% of $CH_4$ remains intact even after 100 s in the 1:1 $CH_4$:$H_2$ mixture at 1100 C. The peak fraction of $C_2H_2$ substantially reduces and becomes on the order of only 10%. With such a reduction, the concentration of $C_2H_2$ becomes comparable to the concentration of $C_2H_4$ which becomes slightly larger and was added to the plots for these temperatures.

We conclude that the inclusion of larger PAH species in the chemical mechanism is important for accurate prediction of the fraction of carbon converted to PAH molecules. Correspondingly, it is also important for determining the fraction of carbon in other species which are predominantly $C_2H_2$. The reduction of $C_2H_2$ concentration due to PAH molecules formation is the strongest at temperatures about 1400C, for both pure and diluted methane. Correspondingly, the effect of the inclusion of larger PAH species in the chemical mechanism is also the strongest at this temperature. The fraction of carbon in $C_2H_2$ drops by factors of 10 and 5 from its peak value in the pure and diluted methane, respectively, when the expanded mechanism up to A37 is used. However, it only decreases by half or less when the mechanism accounting for only small PAH molecules (up to A4) is used. At lower temperature (1300C), final fractions of carbon in $C_2H_2$ and PAH molecules are roughly the same as at 1400C with most of the carbon ending up in PAH molecules, however peak value for $C_2H_2$ is considerably lower. At higher temperatures, both the peak $C_2H_2$ fraction and its final value are considerably higher. Even in the case of pure methane, ~50% of carbon ends up in $C_2H_2$ at 1600C. PAH formation is less thermodynamically favorable at this temperature, especially for small PAH molecules. Most of the material ends up in larger PAH species, up to A37.

To summarize this section, the improved mechanism is suitable for accurate modeling of both stages of the methane pyrolysis process. The first stage is considerably faster than the second one. The inclusion of larger PAH species (up to A37) in the chemical mechanism is important for accurate prediction of the fraction of carbon converted to PAH molecules and, correspondingly, residual fraction of acetylene in the mixture. This mechanism should be beneficial for research aimed at determining optimal pyrolysis conditions with a focus on either efficient conversion of methane to PAH species of particular size and structure or possibly suppressing PAH formation. This might be desirable, e.g., to maximize $C_2H_2$ production or CNT synthesis.

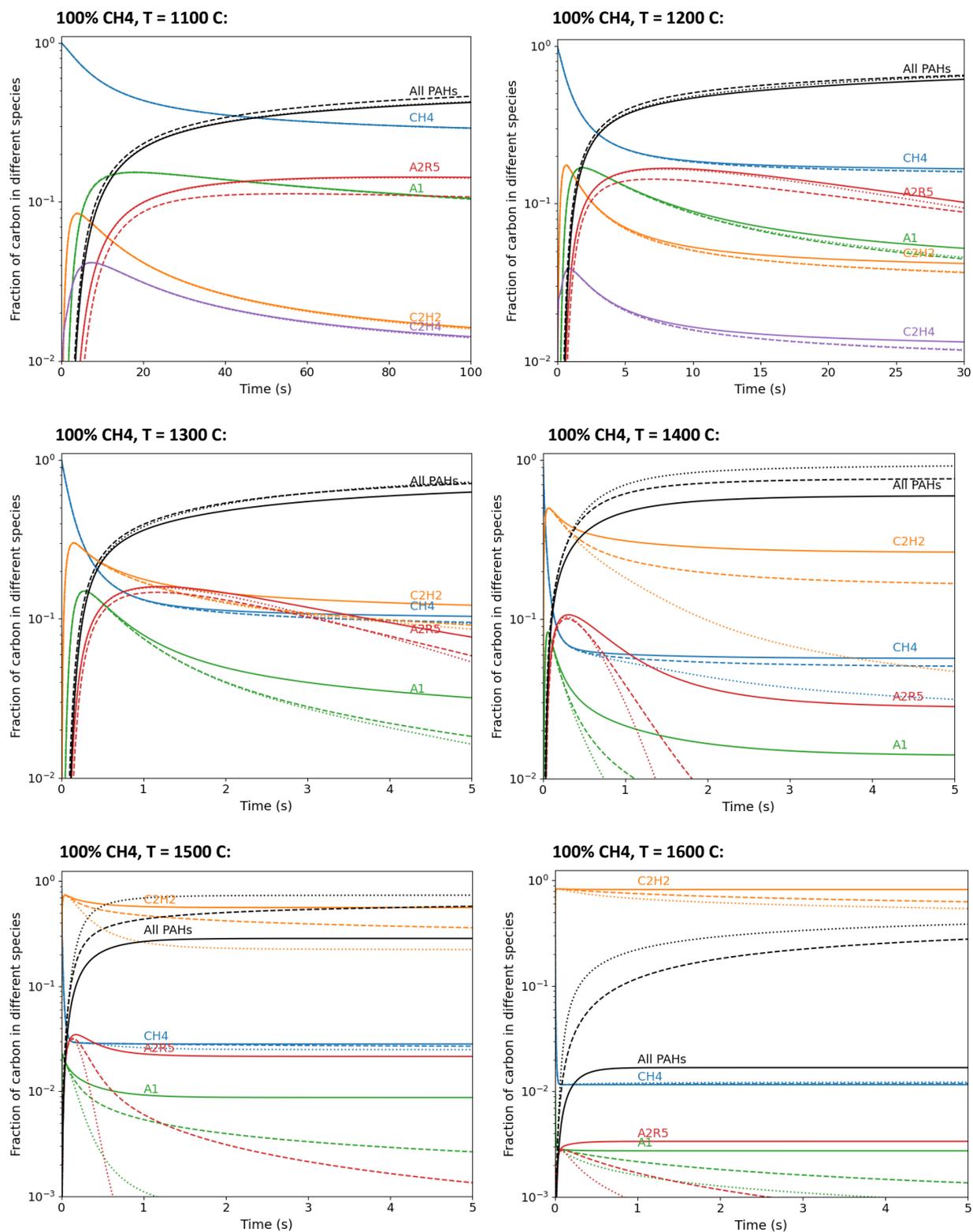

Fig. 9. Modeling results for the evolution of pure methane under different temperatures: fraction of carbon in selected species. Solid lines – ABF[1] mechanism expanded with selected reactions from the Tao mechanism[2], dashed lines – ABF mechanism additionally expanded up to ovalene (A10), dotted lines – ABF mechanism additionally expanded up to A37.

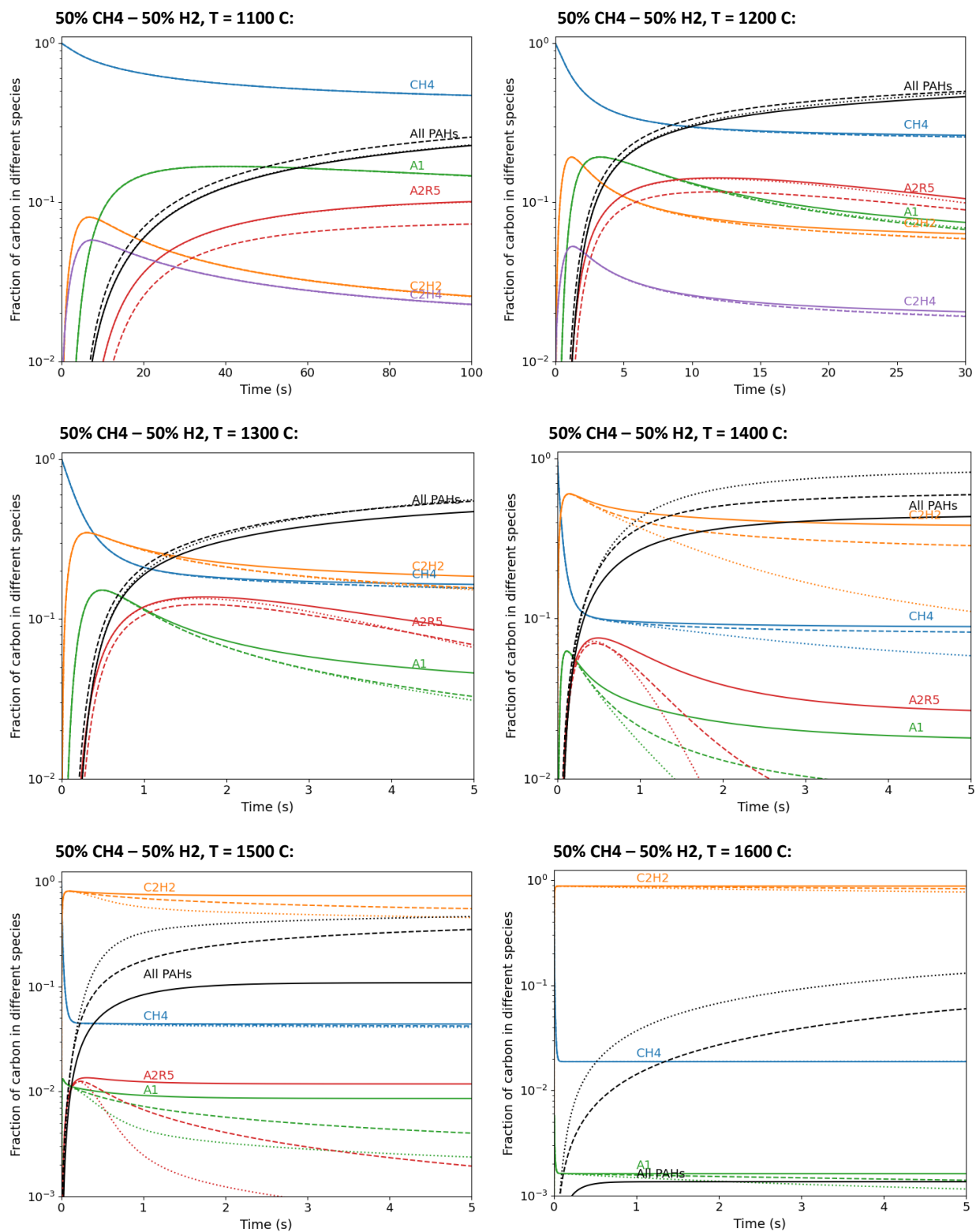

Fig. 10. Modeling results for the evolution of 1:1 methane: hydrogen mixture under different temperatures: fraction of carbon in selected species.

## 5. Summary

A compact and accurate chemical mechanism has been constructed that is capable of modeling both stages of methane pyrolysis process: the fast initial conversion of $CH_4$ to $C_2H_2$ and the slow conversion of $C_2H_2$ to PAH molecules. This mechanism can be used for models describing formation and growth of carbon nanostructures such as CB and soot particles. The mechanism is based on the ABF[1] mechanism for pyrolysis of methane which was expanded with most prominent reaction pathways from the Tao[2] mechanism for small PAH molecules and HACA pathways for larger PAH molecules, up to A37 (37 aromatic rings). The resulting mechanism was validated through comparison to multiple available sets of experimental data for pyrolysis of methane at various conditions and for pyrolysis of acetylene leading to formation of PAH molecules. Good agreement with the experimental data for both processes was obtained. Based on these tests, the proposed mechanism appears to outperform previous mechanisms.

Performance of the proposed mechanism was tested for pyrolysis of methane-rich mixtures under long residence times leading to abundant formation of PAH molecules. It was shown that the inclusion of larger PAH species (up to A37) in the chemical mechanism is important for accurate prediction of the fraction of carbon converted to PAH molecules and, correspondingly, the residual fraction of acetylene in the mixture. This mechanism is designed to model methane pyrolysis with the aim to either create PAH species of particular size and structure or to suppress PAH formation as would be needed to maximize $C_2H_2$ production or CNT synthesis.

**Acknowledgements**

The authors thank ExxonMobil and Princeton University for funding this project.

**Appendix 1. Benchmarking of the mechanism and the plug flow model implemented in Cantera (by comparing to Tao's modeling results).**

We implemented the Tao[2] mechanism in Cantera chemical kinetic solver[62] and reproduced numerical simulations for the conditions of experiments[42] using a plug-flow reactor model. In the experiments, a tubular flow reactor with a short pre-heating zone and a longer preset temperature zone was used. In Ref. [2], an energy transfer equation was solved in their 1D plug-flow model to accurately describe temperature variation in both preheating zone and preset temperature zone (temperature variation in the latest was due to exothermic PAH growth). The results suggested that the temperature variation could be rather accurately described by a simple piecewise-linear profile where linear temperature growth in the pre-heating zone is followed by constant temperature in the preset temperature zone:

$$T(t) = T_{max} + \Delta T_{preheat} \frac{\max{(t_{preheat} - t, 0)}}{t_{preheat}}$$

which rather closely resembles the temperature profile obtained in Ref. [2] using a 1D heat transfer model. Here, $T_{max}$ is the preset temperature in the heated zone of the tubular reactor, $t$ is the flow time (independent variable), $t_{preheat}$ is the duration of linear heating of the gas before it enters the constant-temperature zone which was determined as $t_{preheat} = 2500K/T_{max}(s)$, $\Delta T_{preheat}$ is temperature variation during the gas heating for which a fixed value of 30 K was used. Residence time within the preset temperature (constant-temperature) zone has an inverse-proportional relation to the temperature to account for thermal expansion of the gas (constant flow rate was used in the experiments).

We implemented Tao's chemical mechanism in the Cantera chemical solver and used this simplified temperature profile in the Cantera's plug-flow model. Modeling results in Cantera have been benchmarked by comparing to the original solution[2]. In Fig. A1.1, the comparison between our modeling and the original results from Ref. [2] is shown for final concentrations of four PAH species after the residence time within the preset temperature zone of $t_{res} = 4552K/T_{max}(s)$ (total modeling time is $t_{total} = t_{res} + t_{preheat}$). Perfect agreement between the results suggests correct implementation of the chemical mechanism in Cantera solver and validity of the simplified temperature profile.

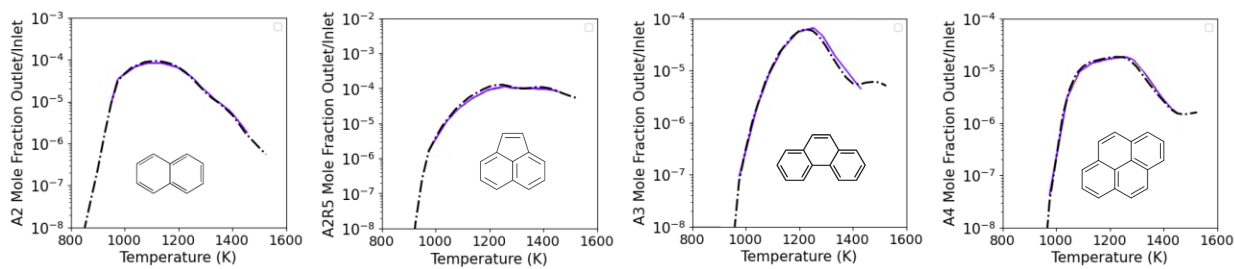

Fig. A1.1. Comparison of final concentrations of four PAH species obtained using the Tao mechanism in the original paper[2] (blue line) and in implementation (dash-dotted black line) for the conditions of experiments[42] at residence time $t_{res} = 4552K/T_{max}(s)$.

**Appendix 2. A list of reactions and rate constants added to the ABF mechanism**

Table 1. Reaction mechanism rate coefficients added to the ABF mechanisms in the Arrhenius form $k = AT^b \exp(-E_a/RT)$ and corresponding literature sources. Units are moles, cm$^3$, seconds, $K$, and calories/mole).

| Reaction | A | b | $E_a$ | References |
|---|---|---|---|---|
| $C_2H_2 + C_2H_2 \Leftrightarrow C_4H_2 + H_2$ | 1.5e+13 | 0.0 | 4.268479e+04 | [48, 63, 64] |
| HCCCCH (+M) $\Leftrightarrow$ $C_4H_2$ + H (+M) | | | | [49, 65] |
| low-P-rate-constant: | 2.0e+15 | 0.0 | 4.0e+04 | |
| high-P-rate-constant: | 1.0e+14 | 0.0 | 4.7e+04 | |
| HCCCCH + $C_2H_2$ $\Leftrightarrow$ | 3.0e+11 | 0.0 | 1.49e+04 | [49, 65] |
| $C_4H_4 + C_2H_2 \Leftrightarrow C_6H_6$ | 4.47e+11 | 0.0 | 3.010177e+04 | [51] |
| A2R5 + H $\Leftrightarrow$ A2R5-3 + $H_2$ | 2.5e+14 | 0.0 | 1.6e+04 | [66] |
| A2R5 + H $\Leftrightarrow$ A2R5-4 + $H_2$ | 3.23e+07 | 2.1 | 1.583688e+04 | [73] |
| A2R5-3 + $C_2H_2$ $\Leftrightarrow$ A2R5E-3 + H | 1.12e+26 | -3.42 | 2.086101e+04 | [67] |
| A2R5-4 + $C_2H_2$ $\Leftrightarrow$ A2R5E-4 + H | 1.12e+26 | -3.42 | 2.086101e+04 | [67] |
| A2R5E-3 + H $\Leftrightarrow$ A2R5E34 + $H_2$ | 3.23e+07 | 2.1 | 1.583688e+04 | [73] |
| A2R5E-4 + H $\Leftrightarrow$ A2R5E45 + $H_2$ | 3.23e+07 | 2.1 | 1.583688e+04 | [73] |
| A2R5E34 + $C_2H_2$ $\Leftrightarrow$ A3LR5* | 1.87e+07 | 1.79 | 3261.03 | [68] |
| A2R5E45 + $C_2H_2$ $\Leftrightarrow$ A3R5-7* | 1.87e+07 | 1.79 | 3261.03 | [68] |
| A3R5-7* + H $\Leftrightarrow$ A3R5 | 5.0e+13 | 0.0 | 0.0 | [67] |
| A3LR5* + H $\Leftrightarrow$ A3LR5 | 5.0e+13 | 0.0 | 0.0 | [67] |
| A3-1 + $C_2H_2$ $\Leftrightarrow$ A3R5 + H | 3.59e+22 | -2.5 | 1.615462e+04 | [67] |
| A3LR5 $\Leftrightarrow$ A3R5 | 8.51e+12 | 0.0, | 6.283626e+04 | [69] |
| $C_4H_5$-2 $\Leftrightarrow$ I-$C_4H_5$ | 1.5e+67 | -16.89 | 5.91002e+04 | [34] |
| $C_4H_5$-2 + H $\Leftrightarrow$ L-$C_3H_3$ + $CH_3$ | 1.0e+14 | 0.0 | 0.0 | [65] |
| $C_4H_4$ + L-$C_3H_3$ $\Leftrightarrow$ C7H7 | 9.25e+11 | -1.265 | -1.4295e+04 | [54, 70] |
| $C_7H_7 + C_2H_2$ $\Leftrightarrow$ $C_9H_8$ + H | 1.0e+11 | 0.0 | 6997.47 | [65] |
| $C_9H_7$ + H $\Leftrightarrow$ $C_9H_8$ | 2.0e+14 | 0.0 | 0.0 | [52] |
| $C_7H_7 + C_9H_7 \Rightarrow$ A4 + $2H_2$ | 2.0e+11 | 0.0 | 2000.0 | [52, 70] |

**Appendix 3. Testing of the original and expanded ABF mechanisms in comparison to other available mechanisms for a broad range of experimental conditions.**

In this section, we thoroughly compare the performance of the original and expanded ABF[1] mechanisms to the Tao's[2] mechanism as well as other mechanisms commonly applied to modeling pyrolysis/combustion of $CH_4$ which include the Fincke[71], Vourliotakis[72], and Smith[73] mechanisms. The latest one is also known as the GRI-3.0 mechanism. We model $CH_4$ pyrolysis for various experimental conditions and compare to available experimental data for various temperature ranges and $CH_4$ dilution ratios 1) moderate dilution with $H_2$ within temperature range 1300 C – 1500 C [55]; 2) pure $CH_4$ at 765 C [56]; 3) strong dilution with He within temperature range 935 C – 1165 C [57]; 4) strong dilution with $N_2$ within temperature range 1000 C – 1200 C [58]. Such a thorough study of ABF mechanism's performance has not been done before.

On Figs. A3.1 and A3.2, modeling results for the conditions of the experiments [55] are presented where $CH_4$ was moderately diluted with $H_2$. Plug-flow model is used to model the tubular flow reactor with the temperature profile assuming linear temperature increase from room temperature to the setpoint reactor temperature within a given time $t_{preheat}$. In the experiments a cold finger was used to quench the reactor which could be moved along the reactor axis in order to control residence time. Because of this feature, experimental data is presented in a convenient format as a function of residence time (we present our modeling results likewise). Fig. A3.1 shows $CH_4$ conversion degree for various reactor set point temperatures at a fixed dilution ratio of $H_2:CH_4 = 2:1$ (molar), whereas Fig. A3.2 shows $CH_4$ conversion degree for various dilution ratios at the reactor temperature of 1300 C. The preheat time was set to 0.003 s for T=1500 C and 0.015 s for other temperatures. As is clear from the figures, the original and expanded ABF mechanisms produce the results that are very close showing that PAH formation is moderate at these temperatures and time frames and has a low effect on overall conversion of methane. Both versions of the ABF mechanism as well as the GRI-3.0 mechanism clearly outperforms other mechanisms producing results that are closer to the experimental data for all temperatures and methane dilution ratios considered in this study. The Tao[2] mechanism overestimates the rate of methane conversion multifold at all temperatures and $CH_4$ dilution ratios. The Vourliotakis[72] mechanism noticeably also overpredicts the rate of methane conversion at all temperatures although to quite as much as the Tao mechanism. The Fincke[71] mechanism considerably underpredicts methane conversion degree at temperatures of 1400 C and below (the lower the temperature the stronger the disagreement with the experimental data).

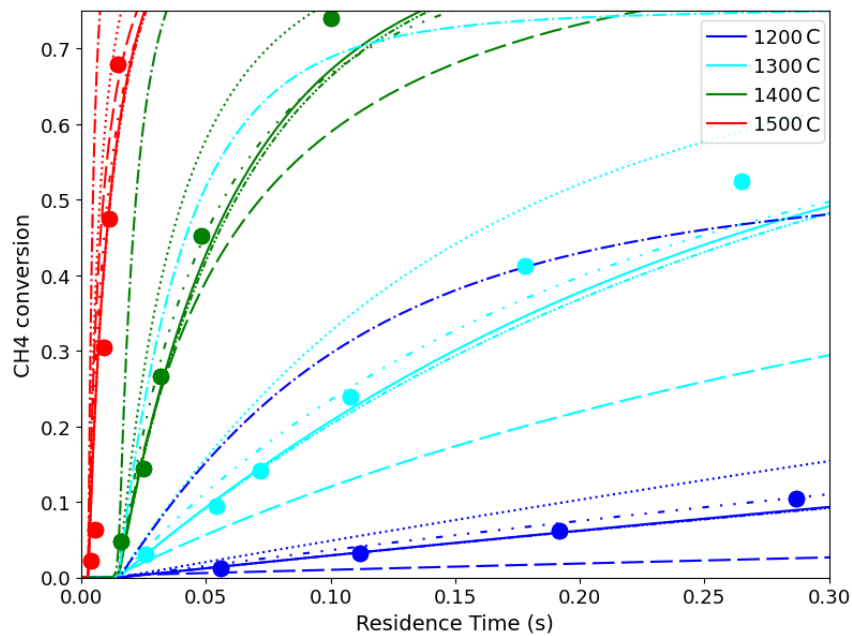

Fig. A3.1. Methane conversion degree as a function of time computed using various chemical mechanisms (lines of different styles) compared to the experimental data[55] (circles) for various reactor temperatures and initial methane dilution ratio of $H_2:CH_4 = 2:1$. ▬•▬••  ABF mechanism (original); ▬▬▬ ABF mechanism added with Tao's reactions and expanded with larger PAH molecules, ▬•▬ Tao mechanism (2019), ••••• Vourliotakis mechanism (2011), ▬ ▬ Fincke mechanism (2002), ▬ •• ▬ GRI-3.0.

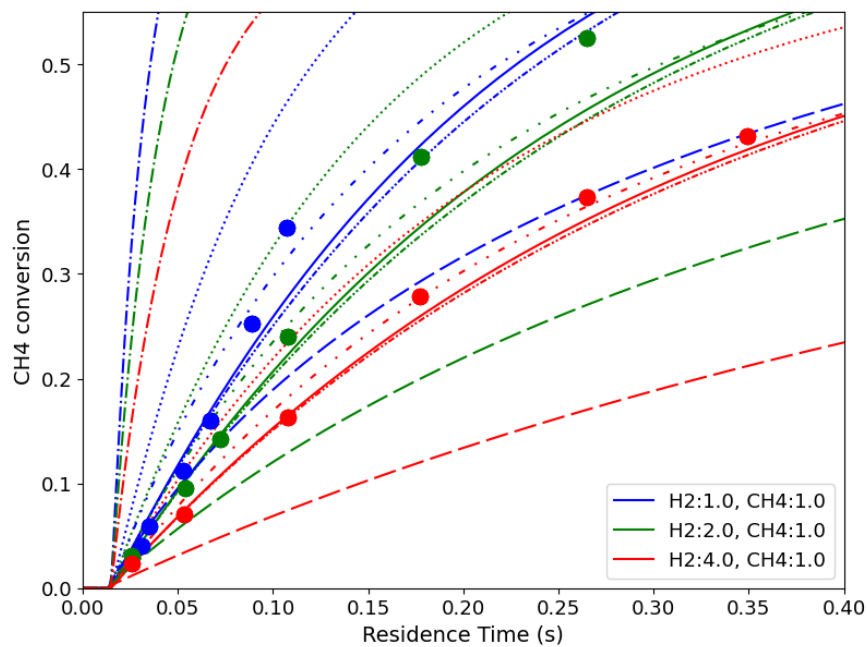

Fig. A3.2. Methane conversion degree as a function of time computed using various chemical mechanisms (lines of different styles, as on Fig. A3.1) compared to the experimental data[55] (circles) for the reactor temperature of 1300 C and various methane dilution ratios.

Modeling results for the conditions of the experiments [56] are presented on Fig. A3.3. In these experiments, pure methane was decomposed in a static reactor (a heated reservoir with no flow) at sub-atmospheric pressure of 441 torr and temperature 1038 K (765 C). Due to low temperatures, the process is slow enabling accurate measurements of the evolution of $H_2$ concentration released from $CH_4$ decomposition. 0D (perfectly stirred) unsteady model was used for this type of reactor. Modeling results from both the original and the improved versions of the ABF mechanism show excellent agreement with the experimental data (the results from the two mechanisms are identical) as well as the Younessi mechanism[25] for which the modeling results were adopted from literature because the mechanism itself is not available. The results of other mechanisms deviate from the experimental values. The Vourliotakis[72] mechanism slightly underpredicts the $H_2$ concentration while other mechanisms considerably overpredict it with various degrees of deviation. The predictions of the GRI-3.0 mechanism and the Fincke[71] mechanism are off by roughly factors of 2 and 3 respectively while the Tao mechanism[2] is off by more than an order of magnitude with H2 fraction growing almost vertically.

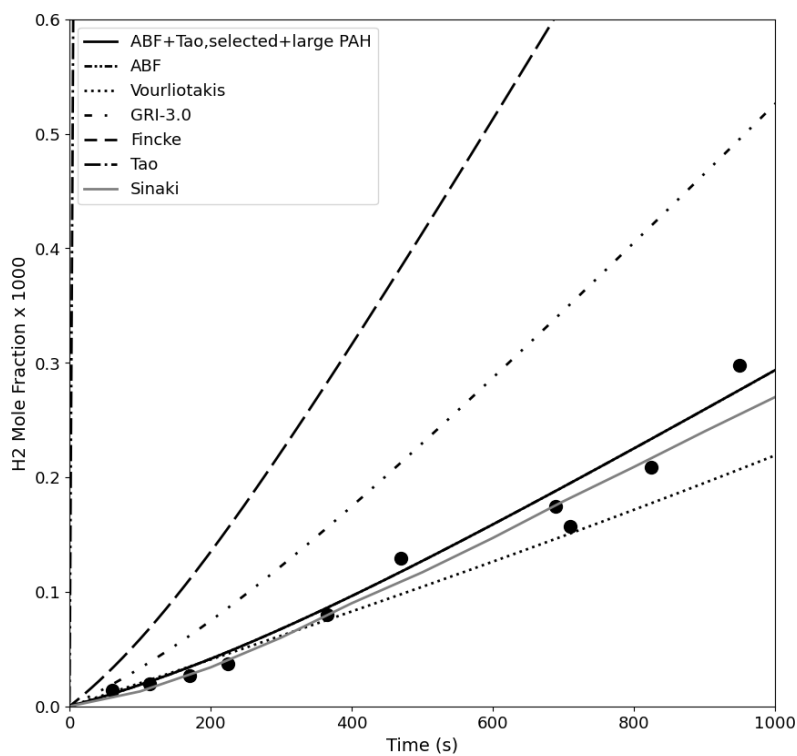

Fig. A3.3. Mole fraction of H2 as a function of time computed using various chemical compared to the experimental data[56] (circles) for the reactor temperature of 1300 C and various methane dilution ratios. Line style denote different chemical mechanisms (same as on Fig. A3.1, except new grey line was added corresponding to the mechanism[25]).

Comparison of the modeling results to the experimental data from Ref. [57] is shown on Fig. A3.4. In these experiments, methane highly diluted in helium at a molar ratio of $CH_4$:He = 9.2:90.8. He passed through a tubular flow reactor with a residence time of 7550K/T (s) at atmospheric pressure conditions. Residual fraction of methane at the outlet of the reactor is plotted on Fig. A3.4 as a function of the reactor temperature. As for the previous experimental conditions, here the results of both versions of the ABF mechanisms as well as the Younessi mechanism[25] are in a very good agreement with the experimental data. The GRI-3.0 mechanism also produces accurate results. The Fincke[71] and Vourliotakis[72] mechanisms underestimate the final $CH_4$ mole fraction (overestimate the $CH_4$ conversion degree) slightly, while the Tao[2] mechanism overestimates the $CH_4$ conversion degree considerably as it did for other experimental conditions.

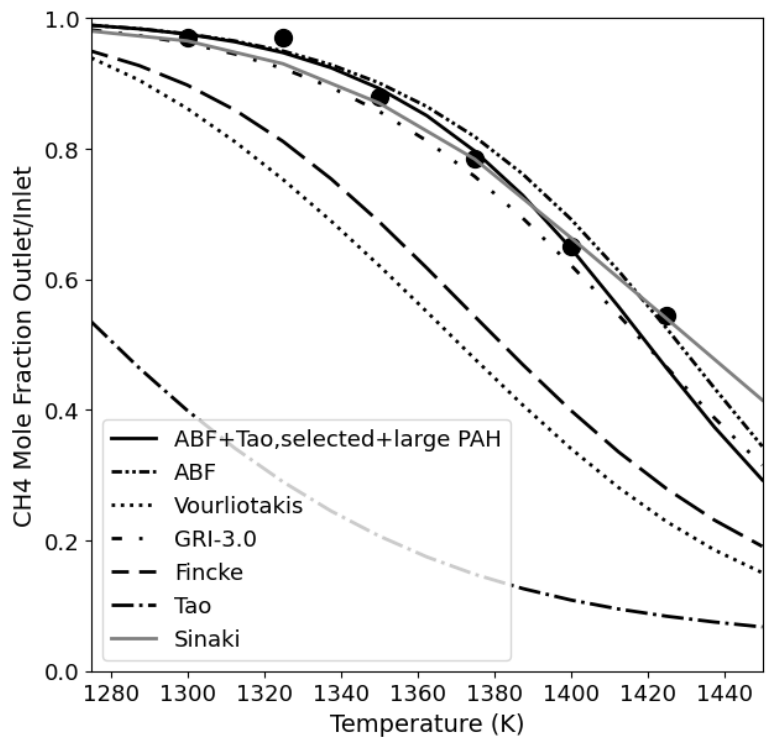

Fig. A3.4. Final mole fraction of $CH_4$ over is initial fraction in a $CH_4$-He mixture after residence time of 7550 K/T (s) as a function of temperature computed using various chemical mechanisms (lines of different styles, as in Fig. A3.1) compared to the experimental data[57] (circles).

Comparison of the modeling results to the experimental data from Ref. [58] is shown in Fig. A3.5. In these experiments, methane highly diluted in helium at a molar ratio of $CH_4:N_2$ = 10:90 passed through a tubular flow reactor with a residence time of 4550K/T (s) at atmospheric pressure conditions. The Vourliotakis[72] and Finke[71] mechanisms perform slightly better than other mechanisms in this case. However, the ABF mechanism (both the original and expanded versions) still exhibits reasonably good agreement with the experimental data. The Tao mechanism, as previously, overpredicts decomposition of $CH_4$ and formation of $H_2$.

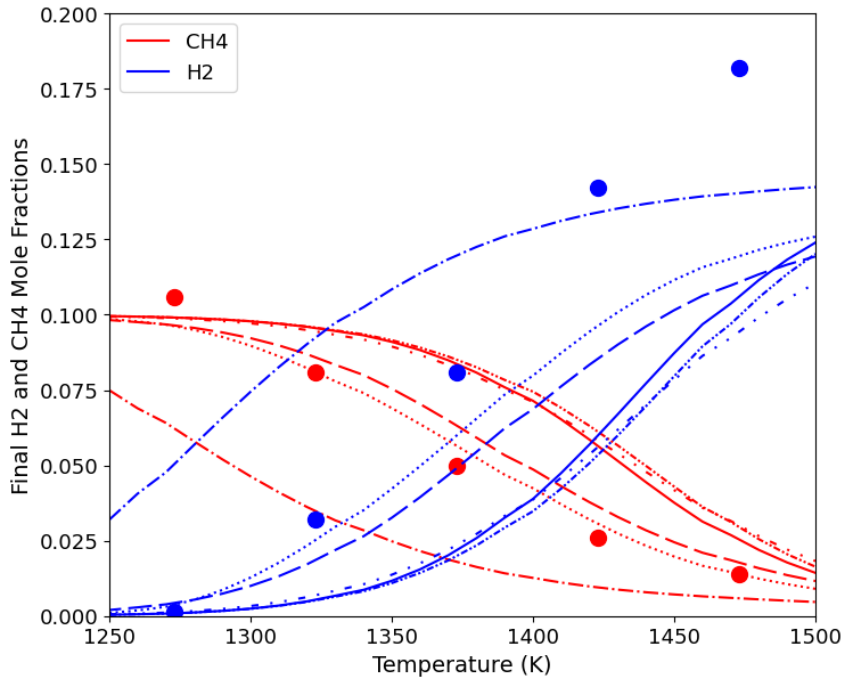

Fig. A3.5. Final CH$_4$ and C$_2$H$_2$ mole fractions in a CH$_4$-N$_2$ mixture after residence time of 4550K/T (s) as functions of the reactor temperature computed using various chemical mechanisms (lines of different styles, as in Fig. A3.1) compared to the experimental data[58] (circles).

To summarize this appendix section, the expanded version of the ABF mechanism has been constructed using selected (most prominent) reaction pathways from the Tao mechanism and HACA pathways for larger PAH species up to A37. This is a compact mechanism, it contains ~200 chemical reactions. This mechanism has been thoroughly tested by comparing to available experimental data and performance of other mechanisms for pyrolysis of methane. It has been shown that the original and expanded ABF mechanisms produce close results for transformation of methane into acetylene. Both mechanisms outperform other mechanisms based on the totality of the tests. One or another of the tested mechanisms may perform marginally better that the ABF mechanisms in one particular set of conditions but fail for other conditions. On the other hand, the ABF mechanisms perform robustly well in all conditions considered.

**Appendix 4. Modeling results for PAH growth from acetylene under expanded residence time**

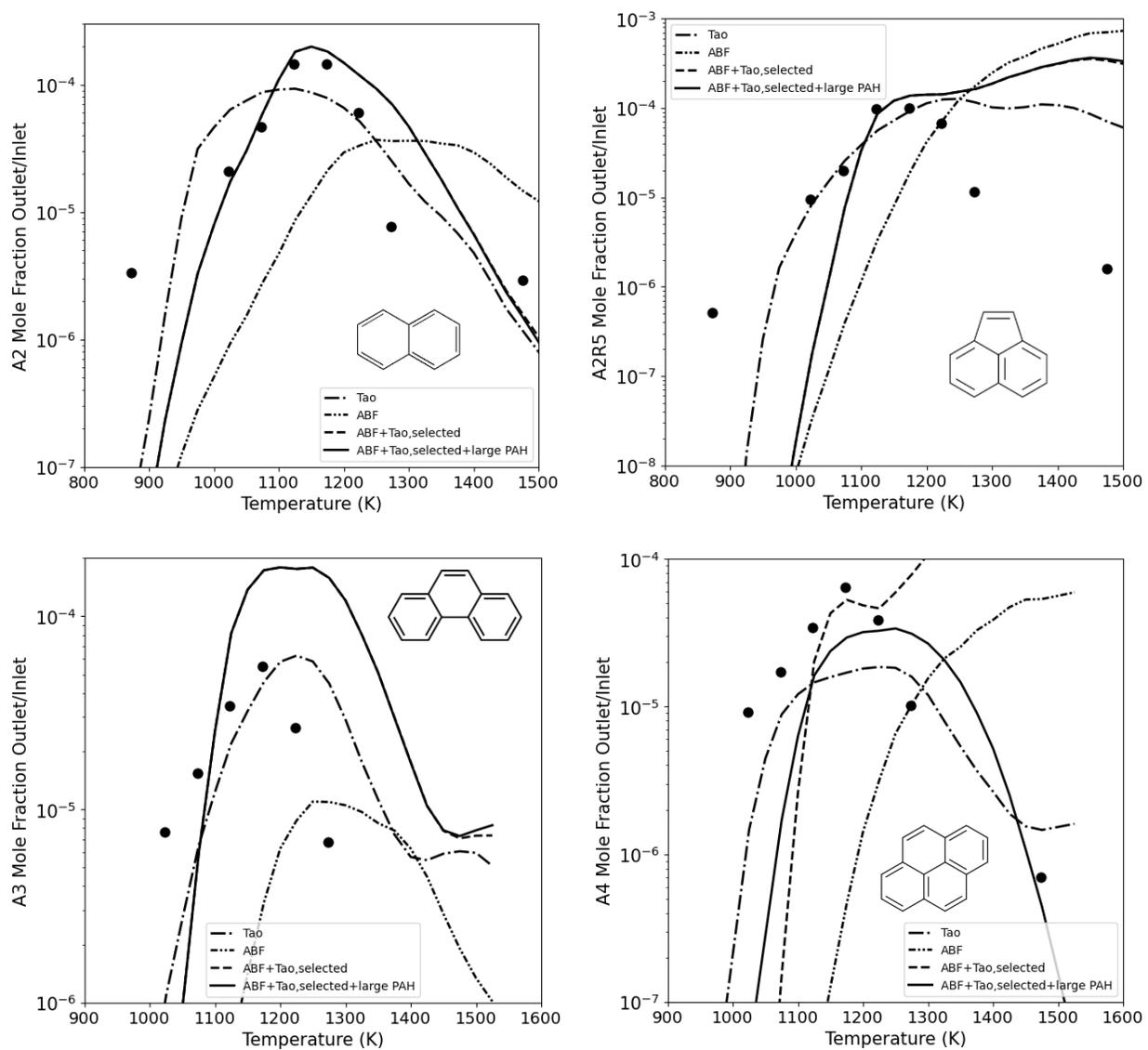

Fig. A4.1. Computed final concentrations of various PAH species at an outlet of a constant-temperature flow from the experiments[42] after the residence time of 4552 K/T (s), i.e., approx. ~4s. Line styles and markers are the same as on Fig. 7.

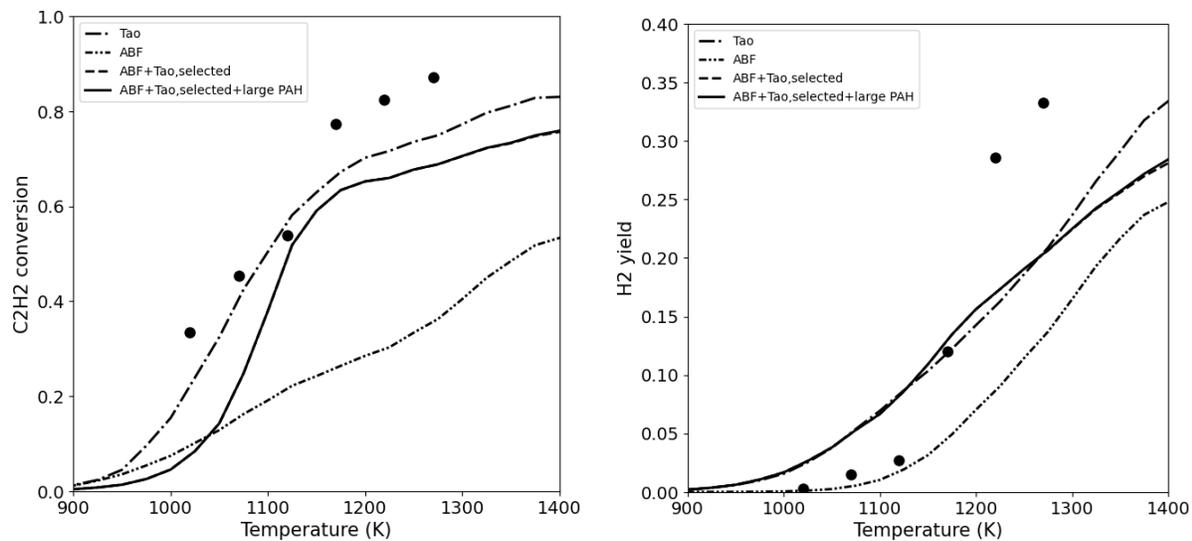

Fig. A4.2. $C_2H_2$ conversion degree and $H_2$ yield after the residence time of 4552 K/T (s). Line styles and markers have the same meaning as on Fig. 7.